\documentclass[fleqn]{article}
\usepackage{epsfig}\parskip 5pt plus 1pt
\usepackage{graphicx}
\usepackage{rotating,amsmath,amssymb,xspace}
\newcommand{\be}{\begin{equation}}
\newcommand{\ee}{\end{equation}}
\newcommand{\bea}{\begin{eqnarray}}
\newcommand{\eea}{\end{eqnarray}}

\def\simge{\mathrel{%
   \rlap{\raise 0.511ex \hbox{$>$}}{\lower 0.511ex \hbox{$\sim$}}}}
\def\simle{\mathrel{
   \rlap{\raise 0.511ex \hbox{$<$}}{\lower 0.511ex \hbox{$\sim$}}}}

\newcommand{\etal}{{\it et al.}}

\def\nue{\ensuremath{\nu_{e}}\xspace}
\def\nubare{\ensuremath{\overline{\nu}_{e}}\xspace}

\def\numu{\ensuremath{\nu_{\mu}}\xspace}
\def\nubarmu{\ensuremath{\overline{\nu}_{\mu}}\xspace}

\def\nutau{\ensuremath{\nu_{\tau}\xspace}}

\newcommand{\numunutau}{\ensuremath{\numu \rightarrow \nutau}\xspace}

\newcommand{\nubarmunubare}{\ensuremath{\overline{\nu}_\mu \rightarrow \overline{\nu}_e}\xspace}
\newcommand{\dmot}{\ensuremath{\Delta m^2_{12}\xspace}}
\newcommand{\dmtt}{\ensuremath{\Delta m^2_{23} \xspace}}

\newcommand{\He}{\ensuremath{^6{\mathrm{He}}}\xspace}
\newcommand{\Ne}{\ensuremath{^{18}{\mathrm{Ne}}}\xspace}
\def\Li8{\ensuremath{^8{\mathrm{Li}}}}
\def\B8{\ensuremath{^8{\mathrm{B}}}}

\newcommand{\thetaot}{\ensuremath{\theta_{13}}\xspace}
\newcommand{\thetatt}{\ensuremath{\theta_{23}}\xspace}
\newcommand{\numunue}{\ensuremath{\nu_\mu \rightarrow \nu_e}\xspace}

\newcommand{\BB}{Beta~Beam\xspace}

\begin{document}
\thispagestyle{empty}
\begin{flushright}
{IFT-UAM/CSIC-10-18}\\
{EURONU-WP6-10-16}
\end{flushright}
\vspace*{1cm}
\begin{center}
{\Large{\bf A minimal Beta Beam with high-Q ions to address CP violation in the leptonic sector} }\\
\vspace{.5cm}

P.~Coloma$^{\rm a,b}$, A.~Donini$^{\rm b,c}$, 
P.~Migliozzi$^{\rm d}$, L.~Scotto Lavina$^{\rm e}$, F.~Terranova$^{\rm f}$ \\
\vspace*{1cm}
$^{\rm a}$ Dep. F\'{\i}sica Te\'{o}rica, Universidad Aut\'onoma de
Madrid, 28049 Madrid, Spain \\
$^{\rm b}$ I.F.T., Universidad Aut\'onoma de Madrid/CSIC, 28049 Madrid,
Spain \\
$^{\rm c}$I.F.I.C., Universitat de Valencia/CSIC, 46071 Valencia, Spain
\\
$^{\rm d}$ I.N.F.N., Sez. di Napoli, Napoli, Italy \\
$^{\rm e}$ University of Zurich, Physik-Institut, CH-8057 Zurich,
Switzerland \\
$^{\rm f}$ I.N.F.N., Laboratori Nazionali di Frascati, Frascati (Rome),
Italy \\

\end{center}

\vspace{.3cm}
\begin{abstract}
\noindent
In this paper we consider a Beta Beam setup that tries to leverage at
most existing European facilities: i.e. a setup that takes advantage of
facilities at CERN to boost high-$Q$ ions ($^8$Li and $^8$B)
aiming at a far detector located at $L = 732$ Km in the Gran Sasso
Underground Laboratory.  The average neutrino energy for $^8$Li and
$^8$B ions boosted at $\gamma \sim 100$ is in the range $E_\nu \in
[1,2]$ GeV, high enough to use a large iron detector of the MINOS type at
the far site.  We perform, then, a study of the neutrino and
antineutrino fluxes needed to measure a CP-violating phase $\delta$ in
a significant part of the parameter space. In particular, for
$\theta_{13} \geq 3^\circ$, if an antineutrino flux of $3 \times
10^{19}$ useful $^8$Li decays per year is achievable, we find that
$\delta$ can be measured in 60\% of the parameter space with $6 \times
10^{18}$ useful $^8$B decays per year.
\end{abstract}

\vspace*{\stretch{2}}
\begin{flushleft}
  \vskip 2cm
{ PACS: 14.60.Pq, 14.60.Lm}
\end{flushleft}

\newpage

\section{Introduction}
\label{sec:intro}

After several years of design and construction, a new generation of
experiments at accelerators (T2K~\cite{T2K}, NOVA~\cite{NOVA}) and
reactors (Double-Chooz~\cite{DoubleChooz}, Daya-Bay~\cite{DayaBay},
RENO~\cite{RENO}) is about to explore {\it subdominant} leptonic
mixing at the atmospheric scale, i.e. oscillations beyond leading
\numunutau transitions at energies and baselines where the oscillation
frequency mainly depends on the mass-squared difference $|\Delta
m^2_{13}| \simeq |\Delta m^2_{23}|$.  At this scale, \numunutau
transitions are driven by the large \thetatt angle ($\thetatt \simeq
45^\circ$~\cite{SchwetzTortolaValle}) while subdominant \numunue
transitions are suppressed by the smallness of the \thetaot mixing
angle between the first and third family. The actual size of \thetaot
is currently unknown and the angle is bounded from above ($\thetaot <
11.5^\circ$), especially by former reactor
data~\cite{chooz,palo_verde} (see, however,
Ref.~\cite{GonzalezGarcia:2007ib} to see some dependence of the upper
bound on $\theta_{13}$ due to the CP-violating phase $\delta$ from
atmospheric data). Three family fits~\cite{Fit th13} and,
particularly, a slight tension between the SNO and Kamland
data~\cite{sno_new} suggest, however, that \thetaot might be close to current
limits. The above-mentioned experiments will be able to probe
values of \thetaot down to about 3$^\circ$~\cite{mezzetto_venice} in 3-5
years from now, and so confirm or disprove this hint for a non-vanishing \thetaot. 

Evidence for \numunue transitions at the atmospheric
scale would be a major breakthrough in neutrino physics: since the
ratio $\dmot/|\dmtt|$ is not exceedingly small  ($\simeq 1/30$), a sizable
\thetaot implies that \numunue oscillations at the atmospheric scale
are heavily perturbed by three family interference effects. As a
consequence, precision measurements of \numunue oscillation and its
CP-conjugate \nubarmunubare using artificial sources at long baselines
become an ideal tool to address CP violation in the leptonic
sector~\cite{derujula}.  

Distilling CP-violating effects from the rate
of appearance of \nue and \nubare is a tremendous challenge, clearly
out of reach for the next round of
experiments~\cite{Huber:2009cw}. The design of a further generation of
facilities specifically aimed to probe CP violation in the leptonic
sector, to perform precision measurement of the \thetaot angle and,
possibly, to establish the sign of \dmtt\ through the exploitation of
matter effects has been at the focus of a decade-long study, which was
recently summarized in the ``International Scoping Study of a Future
Neutrino Factory and Superbeam facility''
Report~\cite{ISS_phys,ISS_det,ISS_acc}. Generally speaking, these
facilities require either the construction of underground laboratories
of unprecedented size to host massive low-density detectors - as for
the options based on ``Superbeams''~\cite{ISS_det} - or of a new major
acceleration complex - as for the case of the ``Neutrino
Factories''~\cite{ISS_acc}. The only notably exception to this scheme
pertains to a sub-class of options based on the \BB
concept~\cite{BB-book}, sometimes called ``high-energy Beta Beams''.
Since its inception~\cite{zucchelli}, Beta Beams have been designed
with the aim of leveraging at most existing facilities and, in
particular, the CERN acceleration complex. As explained in
Sec.~\ref{sec:machine} and \ref{sec:minimal}, Beta Beams that are able
to accelerate radioactive ions to high energies and produce multi-GeV
\nue and \nubare allow for the use of high-density detectors, which,
in turn, might be hosted in moderate-size underground laboratories.
For a CERN-based \BB, the natural option to host the far detector is
a laboratory located at a distance $O(600-700)$ Km from the neutrino source.
The facility that {\it exploits at most existing European infrastructures}, 
as discussed in Sec.~\ref{sec:minimal}, is a multi-GeV \BB based on the CERN-SPS
accelerator pointing to a massive, high-density detector located in
one of the experimental halls of the Gran Sasso laboratories.
The next cheapest alternative could be represented by the Canfranc Underground Laboratories in Spain, 
where some engineering would be however needed (albeit not so impressive as for a Mton class Water \v Cerenkov detector).

The physics performance of this facility and the minimum
requests to the accelerator complex to establish CP violation in the
leptonic sector in case of positive result from T2K, NOVA or the
reactor experiments is at the focus of the present paper
(Sec.~\ref{sec:detector} and \ref{sec:performance}). Beside the huge
practical interest of exploiting in an optimal manner all European
facilities without additional infrastructure
investment~\cite{Battiston:2009ux}, this detailed assessment is
particularly relevant at present times: since 2009, machine studies
for the \BB are concentrated on facilities that accelerate ions with
Q-values larger than originally envisioned ($Q \sim 13$~MeV for $^8$Li and $^8$B, to be compared with
$Q \sim 3$~MeV for the ions considered in the original design, $^6$He and $^{18}$Ne) 
using the existing SPS machine~\cite{euronu,D62009}. This option
ideally fits the ``minimal'' scheme mentioned above provided that
neutrinos are pointed toward the underground halls of LNGS. Other
options either based on low density detectors and/or on new terminal
boosters at larger energies than the CERN-SPS have also been studied
in literature: for details, we refer the reader to Ref.~\cite{BB-book} and,
in particular, to Refs.~\cite{Bouchez:2003fy,Donini:2004hu,Donini:2004iv} for low-Q ions
accelerated by the SPS,
Refs.~\cite{Burguet-Castell:2003vv,Burguet-Castell:2005va,veryhigh_BB,
Huber:2005jk,Donini:2006tt,Donini:2007qt} for high-$\gamma$ Beta Beams
(using facilities different from the SPS to accelerate ions) and
Refs.~\cite{Rubbia:2006pi,Rubbia:2006zv,Donini:2006dx,Agarwalla:2006vf,
Agarwalla:2007ai,Coloma:2007nn,Jansson:2007nm,Winter:2008cn,Agarwalla:2008ti,
Agarwalla:2008gf,Meloni:2008it,Choubey:2009ks,FernandezMartinez:2009hb} for high-Q
Beta Beams (either at low-$\gamma$ and high-$\gamma$).

\section{Beta Beams with high-Q ions}
\label{sec:machine}

In order to overtake the intrinsic limitation of \numu beams
originating from the decay-in-flight of pions, novel sources based on
the decay of muons (``Neutrino Factories'', NF~\cite{derujula,NF}) or of
beta-unstable ions (``Beta Beams''~\cite{zucchelli}) have been
proposed. In both cases, the initial flavor exploited to study
subdominant transitions is \nue or \nubare oscillating into \numu and
\nubarmu, respectively\footnote{On the contrary, intense neutrino
beams based on $\pi$ decays (``Superbeams'') study the T-conjugate of
the transition measured by the Neutrino Factories or the Beta Beams,
i.e. \numunue and \nubarmunubare.}.  In particular, in the \BB, the
experimentalist benefits of a nearly ideal knowledge of the flavor and
spectrum of the neutrinos and, contrary to the Neutrino Factory, of
the presence of just one neutrino flavor in the initial state (\nue\
for $\beta^+$ and \nubare for $\beta^-$ unstable isotopes). However,
the choice of available isotopes is rather narrow: the ions employed
in a \BB\ must be produced at high yields to reach sizable neutrino
fluxes; ions cannot decay too early to allow for acceleration and
injection in a dedicated storage ring (from here on called the ``decay
ring'') equipped with straight sections that point toward the far
detector; on the other hand, ions cannot decay too slow so to have a
sizable neutrino flux at the far detector in a short time, or
equivalently, to avoid that partially filled bunches of ions remain in
the decay ring for too many turns complicating the injection of new
bunches.  Besides the difficulties in producing, accelerating and
storing unstable ions, the \BB\ technology suffers from an important
drawback with respect to the Neutrino Factory. Both facilities
envisage a front-end stage where muons or ions are produced and
manipulated to fit the acceptance of a chain of boosters. The boosters
increase progressively the energy of the particles and at the exit of
the last element of the chain (terminal booster) the muons or the ions
are injected and stacked into the decay ring. Now, for any realistic
terminal booster, the NF will produce neutrinos with much larger energy
than the Beta Beam. A further problem of the Beta Beams
facilities is that, due to the $Z/A$ ratio of the injected ions, much
longer rings are needed in order to maintain an intense flux aiming at
the far detector with respect to Neutrino Factories. The typical size
of a racetrack NF ring is $L_{ring} \sim 1500$ m, compared to
$L_{ring} \sim 7000$ m for $\gamma = 100$ Beta Beams storing $^6$He
and $^{18}$Ne ions~\cite{zucchelli}.

With these problems in mind, the original Beta Beam design was conceived
to leverage at most the CERN accelerator complex and profit of the high isotope production
yield reachable by ISOL techniques~\cite{EurisolDS}: in this framework
the natural terminal booster was the SPS, which can accelerate ions up
to a maximum Lorentz $\gamma$ of $\simeq 450 \cdot Z/A$, while the
choice of the ions came down to \He\ and \Ne\, with Q-values of 3.51
and 3.41~MeV, respectively.  As a consequence, the mean energy of the
neutrinos ($\simeq \gamma Q$) does not exceed $\sim$0.5~GeV
($\sim$0.9~GeV) for \He\ (\Ne)~\cite{Burguet-Castell:2005va}. Massive,
low-density detectors are needed to overcome the smallness of the
cross section and the dominance of quasi-elastic scattering at these
energies, so that the detectors at the far location require, like for
Superbeams, the construction of large underground infrastructures, as
the proposed extension of the Modane Laboratories up to
$10^{6}$~m$^3$~\cite{memphys}. As noted first
in Ref.~\cite{Burguet-Castell:2003vv}, working at larger mean energy has a
remarkable impact on the physics performance provided that ion
production and decay rate can be kept at the same level as for the
previous options. Employing neutrinos in the multi-GeV range exhibits
an additional advantage: the oscillation signal (\numu CC events in
the bulk of unoscillated \nue NC and CC interactions) can be observed
and effectively separated from the background in high density,
moderate granularity detectors~\cite{Donini:2006tt} hosted in
underground halls much smaller than the ones envisaged for
Superbeams, such as the Gran Sasso underground laboratory (LNGS).

One option to achieve a larger mean neutrino energy is to use
a new terminal booster of larger rigidity, as the proposed SPS+~\cite{PAF}, 
that would permit to increase the maximum $\gamma$. A higher $\gamma$
would be ideal\footnote{For a discussion of this
option specifically focused on LNGS, see Ref.~\cite{Donini:2006tt}.} since
larger $\gamma$ are beneficial for the flux at the far detector. 
Notice, however, that the decay rate at the storage ring decreases due to the
larger ion lifetime in the lab frame ($\gamma \tau$).
In fact, the neutrino flux per solid angle in a far detector located at a baseline
$L$ from the source, on-axis with respect to the boost direction of the parent
ion is~\cite{Burguet-Castell:2003vv}:

\begin{equation}
\left.{d\Phi \over dS dy}\right|_{\theta\simeq 0} \simeq
{N_\beta \over \pi L^2} {\gamma^2 \over g(y_e)} y^2 (1-y)
\sqrt{(1-y)^2 - y_e^2},
\label{eq:flux}
\end{equation}
where $0 \leq y={E \over 2 \gamma E_0} \leq 1-y_e$, $y_e=m_e/E_0$
and
\begin{equation}
g(y_e)\equiv {1\over 60} \left\{ \sqrt{1-y_e^2} (2-9 y_e^2 - 8
y_e^4) + 15 y_e^4 \log \left[{y_e \over
1-\sqrt{1-y_e^2}}\right]\right\} \, .
\end{equation}
In this formula $E_0 = Q + m_e$ is the electron end-point energy in the
center-of-mass frame of the $\beta$-decay, $m_e$ the electron mass,
$E$ the energy of the final state neutrino in the laboratory frame and
$N_\beta$ the total number of useful ion decays per year. At larger $\gamma$,
the mean neutrino energy increase as $\gamma Q$ and the flux as $\gamma^2$. 
To evaluate the actual advantage
with respect to the SPS-based option, we consider the number of
events at the far location for setups where the detector is always
positioned at the peak of the oscillation maximum ($|\Delta
m_{23}^2|L/4E = \pi/2$).  For a given number of decays per year
$N_\beta$ in the decay ring, the events at the far location are
proportional to the convolution of the flux ($\phi \sim
\gamma^2/L^2$), of the cross section\footnote{In fact, the energy of
the SPS-based \BB is so small that the linear approximation for the
cross section rise is inappropriate and the advantage of the increase
of $\gamma$ is even larger than what accounted
here~\cite{Donini:2006dx}.}  ($\sigma \sim E \sim Q
\gamma$) and of the oscillation probability, times $N_\beta$. If the
facility is operated at the first maximum of the oscillation probability,
then $|\Delta m^2| L/4E= \pi/2$ and, therefore, $L \sim Q
\gamma$. As a result, the number of events is proportional to
\be
N \simeq \frac{N_\beta \ \gamma}{Q}
\label{eq:scaling}
\ee
An increase of $\gamma$ is therefore beneficial, provided that the ion
decay rate $N_{\beta}$ does not drop faster than
$\gamma^{-1}$~\cite{Huber:2005jk,Donini:2008zz}: this is possible in spite of the
relativistic $\gamma \tau$ increase of the lifetime because, in
general,  a higher $\gamma$ allows for a larger number of ions to be
stacked in the decay ring, since the length of a high-$\gamma$ ion bunch
is reduced by a factor $\gamma$ due to Lorentz contraction and thus the occupancy 
of the ring can be maintained fixed. 
In summary, it can be shown that high-$\gamma$ \BB are outstanding tools to
improve our knowledge of leptonic mixing, but they must still overcome two
important technical challenges:

\begin{itemize}
\item they need a terminal booster with larger rigidity than the SPS and appropriate
power to stand the yield obtained at the ion production front-end;
\item they need a decay ring whose curved sections have a rigidity comparable or larger
than of one of the terminal booster (i.e., they need longer rings with respect to low-$\gamma$ \BB).
\end{itemize}

\noindent
On the other hand, none of these challenges have to be faced if the energy increase of
the neutrino is achieved employing isotopes with larger Q-values: for example, 
the SPS can still be used as a booster; the ring needed to store high-Q ions 
can be a bit shorter than the one proposed for low-Q ions (or ions with a moderately higher $\gamma$ can be stored in the same ring). 
Until 2006, however, \BB with ions different from \He\ and \Ne\ was not considered, one of the motivation 
being the scaling of the number of events at the first peak shown in Eq.~\ref{eq:scaling} (see, however, Ref.~\cite{DoniniatHamburg}).

In Ref.~Ê\cite{Autin:2002ms}, the production using ISOL techniques and the acceleration stage using the PS and SPS have been reviewed. 
A new technique to produce low-Z, high-Q ions, however, was proposed in 2006 by C.~Rubbia~\cite{Rubbia:2006pi} and
Y.~Mori~\cite{Mori:2005zz} and specifically adapted to high-Q \BB in
Refs.~\cite{Rubbia:2006pi,Rubbia:2006zv} through the production of \Li8 and
\B8 as \nubare\ and \nue\ sources, respectively.  This technique is a
new application of ionization cooling~\cite{ionization_cooling}
particularly suited for strongly interacting beams. For the case of
the high-Q \BB, intense beams of $^7 \mbox{Li}$ ($^6 \mbox{Li}$) are
stacked into a small storage ring containing a deuterium ($^3
\mbox{He}$) target. Ionization losses in the target ``cool''
transverse betatron oscillations and the mean energy loss along the
longitudinal direction can by compensated by means of RF cavities. The
longitudinal motion of the beam, however, is intrinsically unstable because
faster particles ionize less than average while slower ones ionize
more and, hence, the momentum spread increase exponentially with time.
Such instability is cured locating the target in a point of the ring
that exhibits a chromatic dispersion, i.e. a point where particles are
displaced from the nominal orbit position proportionally to their
momentum spread. If the thickness of the target grows linearly with
the radius of the orbit, faster particles (orbiting at larger radii)
will cross a thicker area and, therefore will loss more energy by
ionization. Similarly, slower particles orbit at lower radii and lose
less energy than average due to the reduced thickness of the
target. A wedge-shaped target, therefore, can restore stability in the
longitudinal motion. This technique allows for the production of \Li8
through $^7$Li + D $\rightarrow $ $^8$Li + p and \B8 through $^6$Li +
$^3$He $\rightarrow $ $^8$B + n by inelastic collisions of
the stacked ions onto the shaped target. 

Either using the ionization cooling technique or standard ISOL methods, a significant $^8$Li 
flux can be produced (for ISOL techniques, the production rate of $^8$Li is 100 times larger than for $^6$He~\cite{Tengblad}).
The ionization cooling technique should guarantee a similar production rate for $^8$B. 
In this case, however, the main problem resides in the extraction and recollection of $^8$B ions:
they react with many elements typically used in targets and ion sources and are therefore 
difficult to manipulate. Eventually, it must be reminded that the $^8$B $\beta$-decay spectrum is affected by several systematics errors that must be 
tamed before using it for a precision experiment (see Ref.~\cite{Winter:2003ac}).

The issue of ion production and extraction from the storage ring~\cite{euronu}, of the effectiveness of
cooling~\cite{Neuffer:2007zzc} and of the yields actually sustainable
by the SPS are at the focus of current accelerator research in \BB. 
It must be shown, in fact, which are the maximal yields that can be
effectively produced and extracted from the source area for a given low-Q or high-Q ion, 
and that the rest of the booster chain can sustain these yields.

\begin{figure}
\begin{center}
\includegraphics[width=\textwidth]{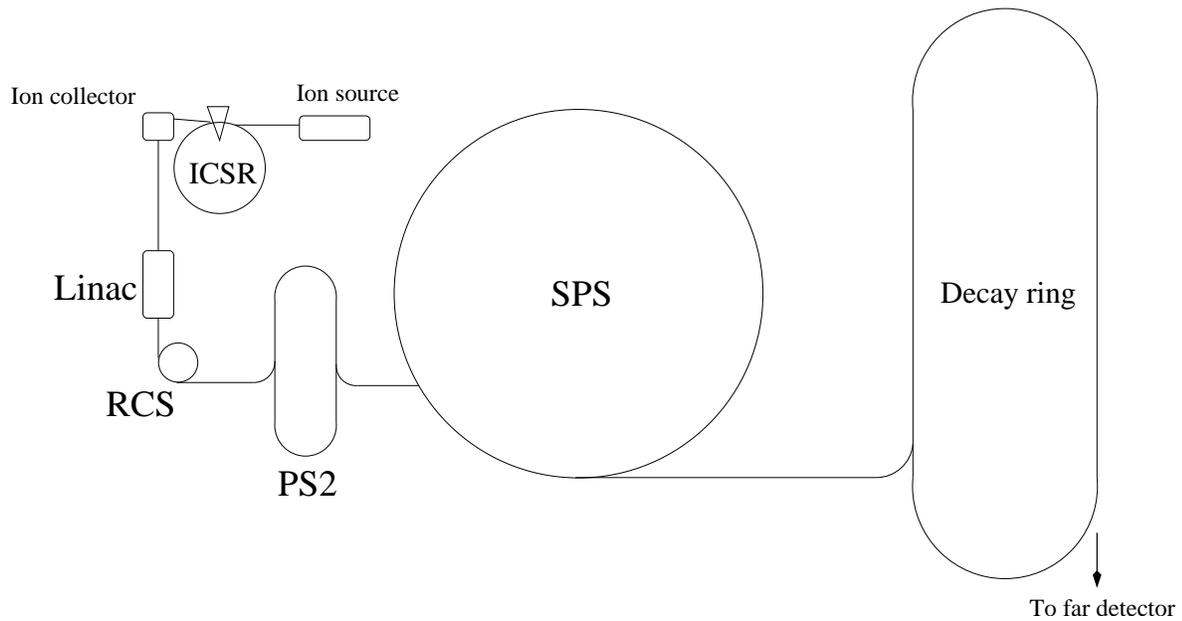}
    \caption{The machine complex for a high-Q Beta Beam in the
EURO$\nu$ scenario. The front-end ion ( $^7$Li and $^6$Li) source
(``Ion source'') is a 20 MeV linac injecting the isotopes into the
ionization cooling storage ring (``ICSR''). Ions produced in the ring
by inelastic interaction with the wedged target are collected,
separated (``Ion collector'') and sent toward a first stage booster
(``Linac''). A dedicated Rapid Cycling Synchrotron (RCS) and the PS2
further accelerate the ions before injection into the terminal booster
(the existing CERN-SPS). Eventually, ions are injected into the decay
ring and stacked. Ions decaying in the straight session of the ring
produce \nue or \nubare pointing toward the far detector.} 
\label{fig:setup}
\end{center}
\end{figure}

In the framework of the EURO$\nu$ Design Study~\cite{euronu} the proposed
\BB (see Fig.~\ref{fig:setup}) exploits the CERN accelerator complex as
it evolves in the forthcoming years to cope with the needs of the LHC. In
particular, the \BB facility does not require the construction of a new
terminal booster, neither as a dedicated machine nor as an ancillary
facility shared with the LHC (as for the case of the SPS+). The only
assumption made on the evolution of the CERN complex is the
replacement  of the Proton Synchrotron with a new
machine (PS2~\cite{PS2}) injecting protons at an energy of 50~GeV into
the SPS. Such replacement is presently envisioned to grant the
reliability of the LHC injection complex and for the luminosity upgrade of the LHC itself.

\section{A facility exploiting at most existing infrastructures}
\label{sec:minimal}

\begin{figure}
\begin{center}
\includegraphics[width=0.8\textwidth]{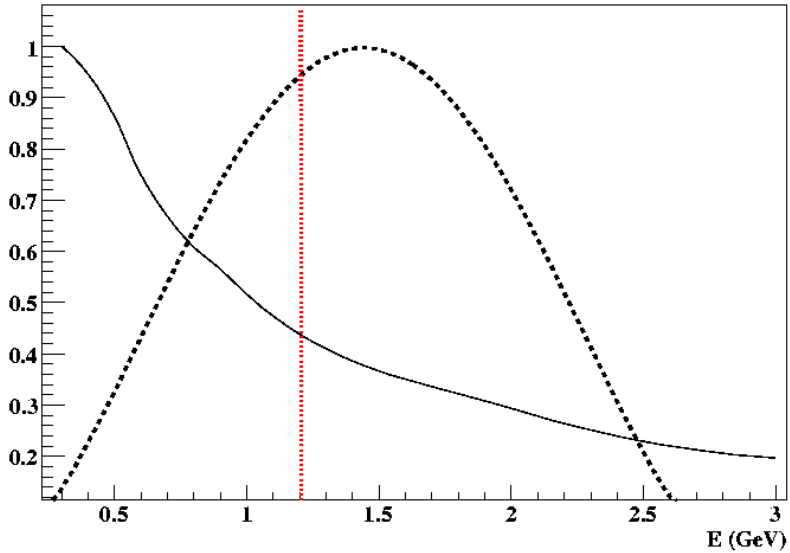}
    \caption{Dashed black line: the flux of neutrinos at the far
detector from the decay of \B8 boosted at $\gamma = 100$ (in a.u.) versus the neutrino energy
(in GeV). The continuous black like represents the ratio between
quasi-elastic and total cross section in \numu CC interactions. The
vertical dotted (red) line shows the muon energy needed to reach a
range in iron larger than 4.6 interaction lengths, corresponding to a
residual contamination of punch-through non-interacting pions lower
than $10^{-2}$. } 
 \label{fig:flux}
\end{center}
\end{figure}

The facility that we consider in this paper does not differ from the
baseline EURO$\nu$~\cite{euronu,FernandezMartinez:2009hb} design, but for 
a high density far detector located in a pre-existing hall at LNGS.  
The rationale behind this choice can be described starting
from Fig.~\ref{fig:flux}.  The dashed (black) line shows the
neutrino energy spectrum (in a.u.)  for \B8 at $\gamma=100$ as a
function of the neutrino energy (in GeV).  In the same plot, the
continuous black line represents the ratio between the quasi-elastic
and total cross section in \numu CC interactions~\cite{lipari}: at the
energy of the neutrinos from high-Q ions boosted by the SPS, such
ratio is heavily suppressed, favoring detectors that are capable to
identify not only quasi-elastic topologies (e.g.  the ``single ring''
events in water Cherenkov detectors) but also deep-inelastic or resonant \numu
interactions (segmented liquid scintillators, liquid argon TPC's or iron calorimeters).
In particular, when the mean muon range exceeds the pion interaction
length, high density detectors (iron calorimeters) can be
employed to separate the \numu CC signal from the bulk
background of \nue NC and \nue CC since hadrons are effectively
filtered by the passive material.  The vertical dotted (red) line in
Fig.~\ref{fig:flux} shows the energy needed by a muon to reach a
range in iron sufficient to filter pions at the $10^{-2}$ level, i.e.
the energy range where signal efficiency for iron calorimeters is
expected to be large (see Sec.~\ref{sec:detector}). As a consequence
the EURO$\nu$ design option is properly suited for the exploitation of
detector technologies based on high density absorbers even with the
present SPS employed as a terminal booster.

Once fixed the detector technology, we must determine the achievable neutrino flux. 
This  is  the most relevant information needed to evaluate the physics performance of this facility, and it depends on
the amount of ``useful decays per year'' of the stored ions, i.e. the amount of ions that decay in the
straight section of the decay ring (see Fig.~\ref{fig:setup}) pointing
to the far detector. This, in turn, depends on the size, geometry and lattice
design of the decay ring (the optimization of which is one of the goals of the EURO$\nu$ accelerator R\&D)
and on the number of ions that can be stored simultaneously into the ring in one year, $N_{ions}$. 

We first discuss the impact of the ring geometry on the neutrino flux.
Independently of the ions chosen, cost and practical considerations
constrain the size of the ring up to about the size of the SPS,
$L_{ring} = 6880$ m.  A decay ring made of relatively small curved
sections (with radius of the curved section $R\sim 300$~m) followed by
long straight sections ($L_{straight}=2500$~m) pointing toward the
detector was considered as a natural geometry for a single far
detector, while for multiple baselines, a triangular geometry might be
more appropriate~\cite{Coloma:2007nn} (see, however,
Ref.~\cite{Choubey:2009ks} and the ISS Physics Final Report,
Ref.~\cite{ISS_phys}).  Decays that provide useful neutrinos are those
occurring in the straight section(s) where neutrinos fly in the
direction of the detector(s).  The useful fraction of decays, called
the ``livetime'' $l = L_{straight}/L_{ring}$ (where $L_{ring}$ is the
total length of the ring), depends on the geometry of the ring: for a
racetrack geometry, $L_{ring}= 2\pi R + 2 L_{straight}$. Once the livetime is eventually fixed, 
then the neutrino flux is given by $N_\beta = l \times N_{ions}$. In the
EURISOL option the livetime was $\sim 36$\% for $L_{ring} = 6880$
m. For fixed straight sections, the livetime depends on the minimum
$R$ that can be achieved, which in turn depends on the gyroradius
$\rho$~\cite{lee} of the ion
\be
\rho = p/|Z|B = \gamma \beta A m_{p}/|Z|B \, ,
\ee
$B$ being the maximum bending field, $m_p$ the proton rest mass and
Z, A the atomic number and weight of the stacked isotope. Since the A/Z
ratio of \Ne and \B8 (1.8 versus 1.6) and of \He and \Li8 (3.0 versus
2.7) are very similar, both pair of ions can circulate in the same
ring geometry, with similar livetime\footnote{Notice that a reduction
of the curved sections of the ring, with a smaller ring size (and,
correspondingly, a reduced cost) and an increased livetime could be
achieved employing superconducting magnets similar to the ones
currently installed for the LHC (8.3~T) . }.  Motivated by these
considerations, in the following we will maintain the same livetimes
as for the EURISOL option, $l = 0.36$.

The number of stored ions $N_{ions}$ depends on a number of factors
related to the production and manipulation of the beam, on which we
can say little (the study of it is part of the EURISOL project goal
for $^6$He and $^{18}$Ne and of the EURO$\nu$ project for $^8$Li and
$^8$B). Once the maximum number of ions that can be collected and
boosted at the desired energy is known, however, they must still be
properly distributed in short bunches inside the storage ring in order
to use time-correlation to reduce the atmospheric neutrino background
at the far detector.  The required duty cycle is very demanding:
previous analysis showed that for $^6$He and $^{18}$Ne ions boosted at
$\gamma \sim 100$ the decaying ions must be accumulated in very small
bunches occupying just a very small fraction of the storage ring. It
has been shown in Refs.~\cite{Mezzetto:2005ae,FernandezMartinez:2009hb}
that a $10^{-3}$ suppression of the atmospheric background (i.e., a
$10^{-3}$ duty cycle) is needed in order to achieve a good sensitivity
to physics observables. This puts a strong constraint on the
manipulation of the beam that can result in a reduction of the
ultimate neutrino flux (being the storage ring not used at its maximum
capacity). In other words, although a given number $N$ of ions can be
produced and accelerated at the desired boost, only a fraction of them
can be actually injected into the storage ring if we want to attain
the required atmospheric background suppression. 

It has been suggested in several papers that increasing the neutrino
energy allow for less demanding duty cycles (see, for example,
Refs.~\cite{Burguet-Castell:2003vv,Burguet-Castell:2005va} and
\cite{Coloma:2007nn}), that in turn permit to store more ions into the
storage ring (see Ref.~\cite{Benedikt:2006hk}).  The net result is
that either a larger neutrino flux aims at the far detector or,
alternatively, a lesser technological effort would be needed to
achieve the desired flux. Unfortunately, in
Ref.~\cite{FernandezMartinez:2009hb} it was shown that it is not
possible to relax the duty cycle too much for $^8$Li and $^8$B ions
boosted at $\gamma = 100$ with respect to $^6$He and $^{18}$Ne boosted
at the same $\gamma$ without losing sensitivity to the physics
observables, assuming that in both setups an $O (1)$ Mton class water
\v Cerenkov is used as far detector In fact, most considerations hold
for the present setup, as well. We will assume in the rest of the
paper a suppression factor in the ballpark of $10^{-3}$ , therefore
neglecting background from off-time atmospheric neutrinos.

The nominal fluxes proposed in the EURISOL project for $^{18}$Ne and
$^6$He are $ 1.1 \times 10^{18}$ \Ne and $ 2.9 \times 10^{18}$ useful
decays per year, respectively.  Preliminary studies show that the nominal flux
is at hand for $^6$He ions (the estimations actually
yield a flux somewhat larger, of $3.18 \times 10^{18}$ useful decays
per year). In the case of $^{18}$Ne, on the other hand, the production
of an intense flux is much more challenging and the present estimates
fall two orders of magnitude short of the mark, yielding a flux of
$4.6 \times 10^{16}$ useful decays per year (see
Ref.~\cite{betabeamwebpage}). As discussed in Sec.~\ref{sec:machine},
significantly larger fluxes are expected from the use of ionization
cooling for high-Q isotopes (although it is not clear if $^8$B ions
can be recollected in huge numbers). For this reason, in the rest of
the paper, we will study the performance of our setup as a function of
the achievable neutrino and antineutrino flux, $F$ and $\bar F$, with
respect to a nominal flux $F_0$, for  $F_0 = 3 \times 10^{18}$ useful decays per
year for both \B8 and \Li8.

\section{Detector simulation}
\label{sec:detector}

The description of a massive iron detector capable to exploit a
high-energy Beta Beam and the efficiencies and background
calculations, have been detailed in Ref. \cite{Donini:2006tt}.
Here, the detector consisted of a sandwich of 4~cm
iron slabs interleaved with glass RPC's to reach an overall
mass of 40~kton. The RPC are housed in a 2~cm gap while the active element
is a 2~mm gas-filled gap, whose signal is digitally read-out by $2 \times 2$
~cm$^2$ pads.  A full GEANT3~\cite{GEANT3} simulation of
this geometry has been implemented along the lines discussed
in Ref.~\cite{tdf}, including a coarse description of the RPC materials and
an approximate description of the digitization process. As detailed
in Ref.~\cite{tdf}, the accuracy of this simulation was validated by
comparing its predictions to existing data collected with a small
prototype exposed to a pion beam of energy from 2 GeV to 10
GeV~\cite{gustavino}. 

\begin{figure*}[tbph]
\centering
\includegraphics[width=75mm]{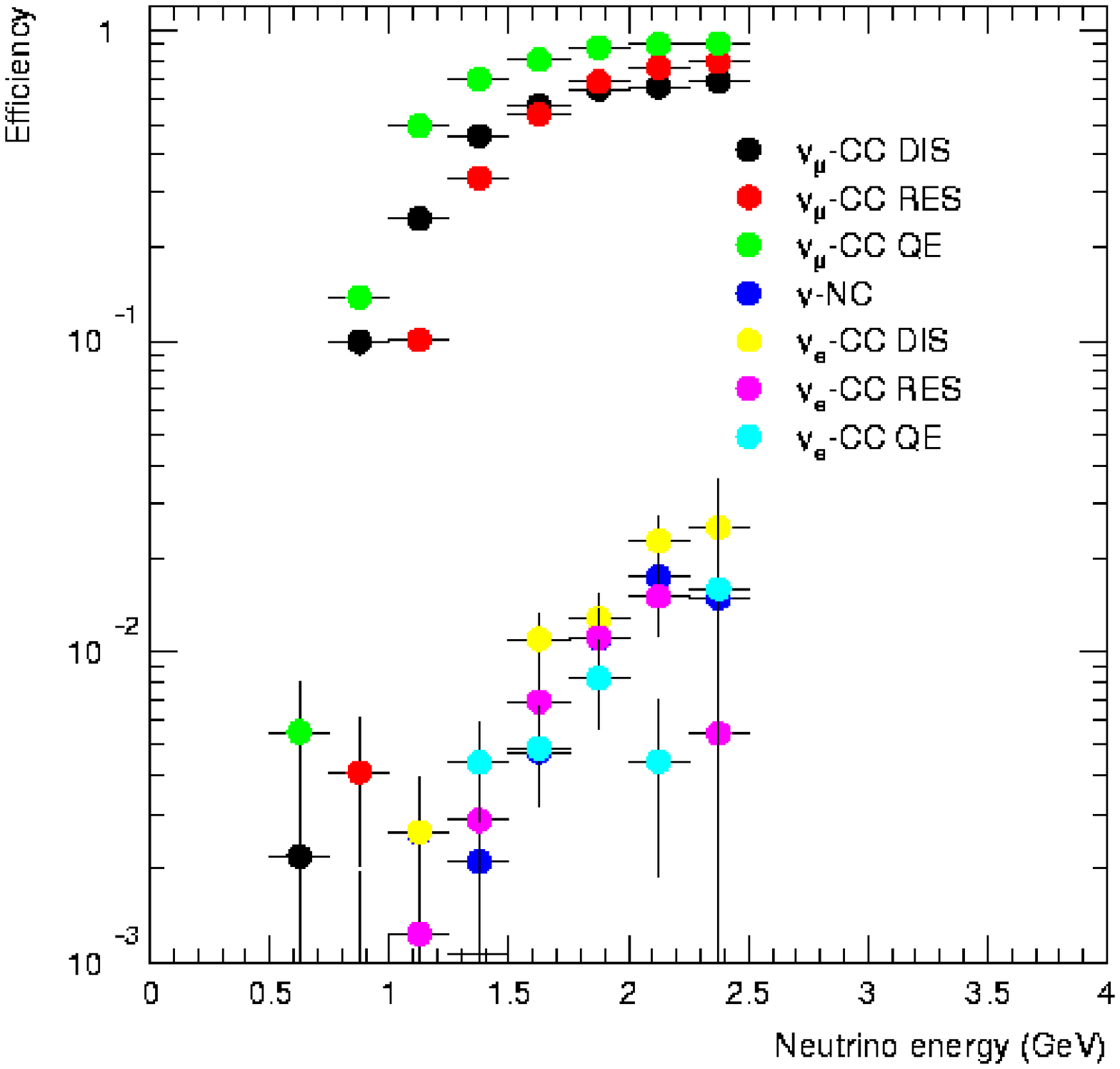}\includegraphics[width=75mm]{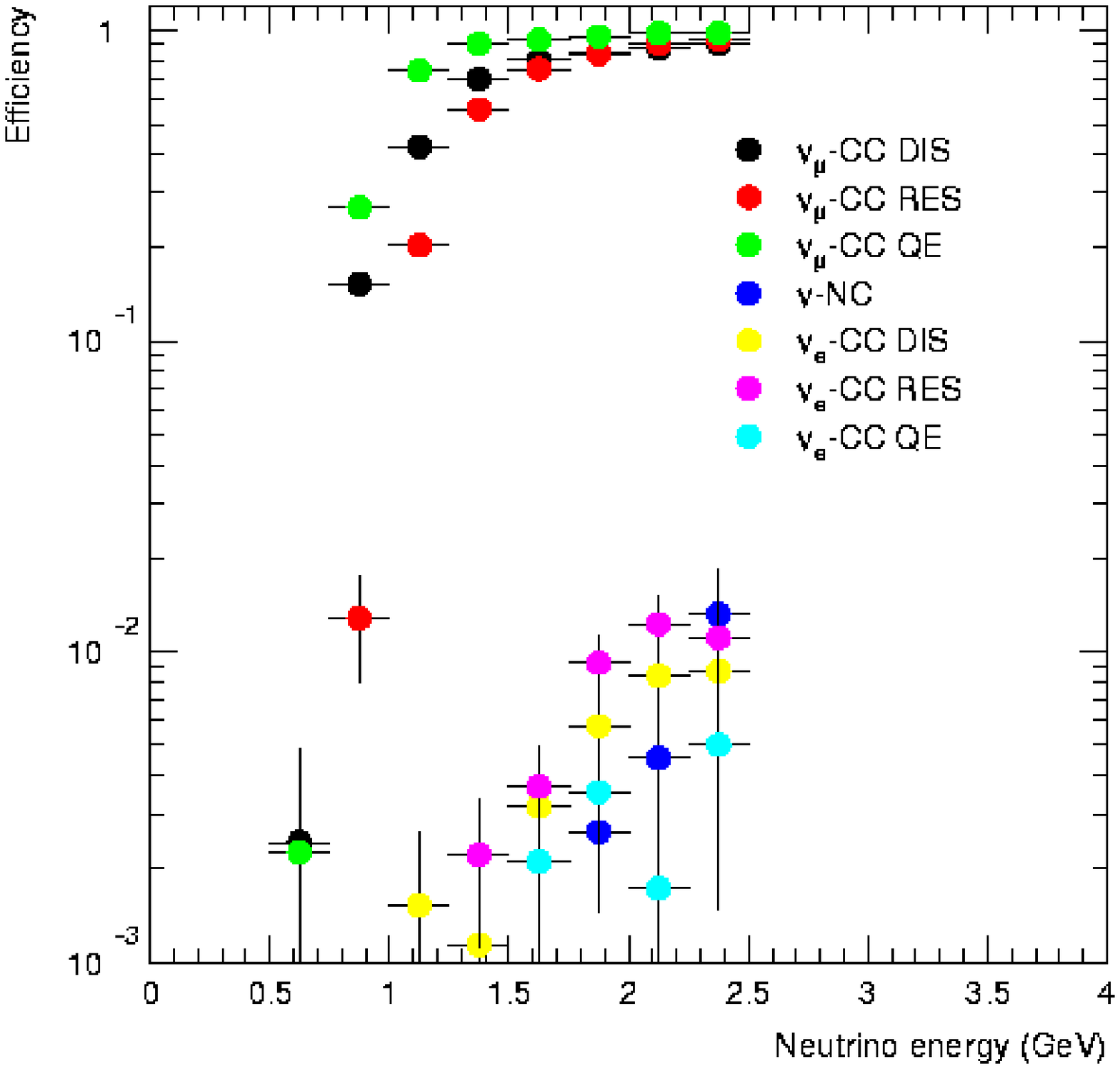}
\caption{Efficiencies for the signal ($\nu_\mu$ and $\bar{\nu}_\mu$
charged-current interactions) to be identified as CC-like event and
for the background ($\nu_e$ and $\bar{\nu}_e$ interactions, and
$\nu_\mu$ and $\bar{\nu}_\mu$ neutral-current interactions) to be
mis-identified as a CC-like events.}
  \label{fig:efficiency}
\end{figure*}

The variables that are used for event classification are purely
inclusive: the total number of hits and event length expressed in
terms of number of crossed iron layers. Since in the high-Q
configuration considered in Sec.~\ref{sec:minimal} the mean neutrino
and antineutrino energies are both 1.5~GeV, we employ the same
selection both for $\nu_\mu$ CC and $\bar{\nu}_\mu$ CC events: an
interaction is classified as a $\nu_\mu$ ($\bar{\nu}_\mu$) CC if both
the event length and the total number of hits in the detector are
larger than 12.  The efficiency for identifying a neutrino CC
interaction averaged out over the whole spectrum is $\sim
60$\%. Conversely, the probability for the background to be identified
as a CC-like event is slightly less than 1\%.  Efficiencies and
background contaminations as a function of the neutrino energy are
shown in Fig.~\ref{fig:efficiency} (see also
Ref. \cite{Donini:2006tt}).

It is worth noting that for larger energies the performance of the
detector is comparable with the one proposed for the Neutrino
Factory, the mean neutrino energy at NF being much larger ($\sim
30$~GeV) than for a high-Q Beta Beam.  The main difference is due to
the fact that here the magnetization of the iron is not mandatory
since Beta Beams are pure sources of $\nu_e$ and the
identification of the muon charge at the final state is immaterial.
Charge identification is only beneficial to reduce the background of
punch-through pions while for the NF it is essential to veto ``right
sign'' events originating from $\bar{\nu}_\mu$ produced at the
source~\cite{ISS_det,NF}. The same detector considered here can serve
as a far detector for a Neutrino Factory when the iron is magnetized
by magneto-motive forces comparable to the one envisaged for
MIND~\cite{MIND,nelson}.

Once more, it is interesting to assess the configuration that exploits
at most existing infrastructures, with emphasis on the opportunities
offered by the LNGS experimental Halls. In~\cite{Donini:2006tt} we
considered a setup inspired by the MONOLITH proposal
\cite{monolith}. MONOLITH, in its original design, would have been
installed in the position presently occupied by the OPERA
experiment~\cite{opera}. Such allocation in the Hall C of LNGS limited
the fiducial mass up to 34~kton. In fact, the Hall C is much larger:
it includes the Borexino experiment~\cite{borexino}, the test facility
CTF between Borexino and OPERA and the pseudocumene storage tanks
positioned just in front of OPERA. Therefore, the maximum longitudinal
size that can be allocated is 90 m, assuming the Beta Beam far
detector to be the only user of the Hall. To make the installation
feasible, the horizontal size cannot exceed 14.5 m. Hence, the maximum
mass conceivable in Hall C is $\sim$100 kton. As noted in the
framework of the NF~\cite{ISS_det}, an investment aimed at increasing
the mass of the detector offers a better cost/benefit ratio with
respect to a prolongation of the data taking of the NF or the
Beta Beam. Such benefit is lost if a dedicated underground hall must
be built on purpose to extend the detector inside a deep underground site.

\section{Results}
\label{sec:performance}

In this section we will study the physics performance of the proposed setup in terms of two observables, defined as follows:
\begin{description}
\item[the CP discovery potential:] for a given point in the parameter space, we will say that CP violation can be discovered if we can rule out the no CP violation hypothesis ($\delta=0^\circ$ and $180^\circ$) at $3\sigma$ 1 d.o.f., after marginalizing over all the remaining parameters for both possible hierarchies.

\item[the $\textrm{sgn}(\Delta m_{23}^2)$ reach:] this is defined as the region of the ($\sin^2 2 \theta_{13},\delta$) plane  for which the wrong hierarchy can be eliminated at $3\sigma$. 
Below this value of $\sin^2(2\theta_{13}) $, the predictions for the wrong hierarchy cannot be distinguished from the data corresponding to the right hierarchy, at a statistical significance of $3\sigma$.

\end{description}

Notice that, in both cases, results will be presented as a function of
$\sin^2(2\theta_{13})$. However, as we have already explained in
Sec.~\ref{sec:minimal}, the physics reach of the setup strongly
depends on the achievable fluxes. Therefore, in the next subsections
results will also be presented as a function of the flux ratios
$F/F_0$ and $\bar F/ F_0$, being $F_0 = 3 \times 10^{18}$ useful
decays per year. A 100~kton detector mass is assumed together with a
data taking duration of 5 years in neutrino and 5 year in antineutrino
mode.

\subsection{Sensitivity to the CP-violating phase}
\label{subsec:CP}

\begin{figure}
\begin{center}
{\includegraphics[width=0.6\textwidth]{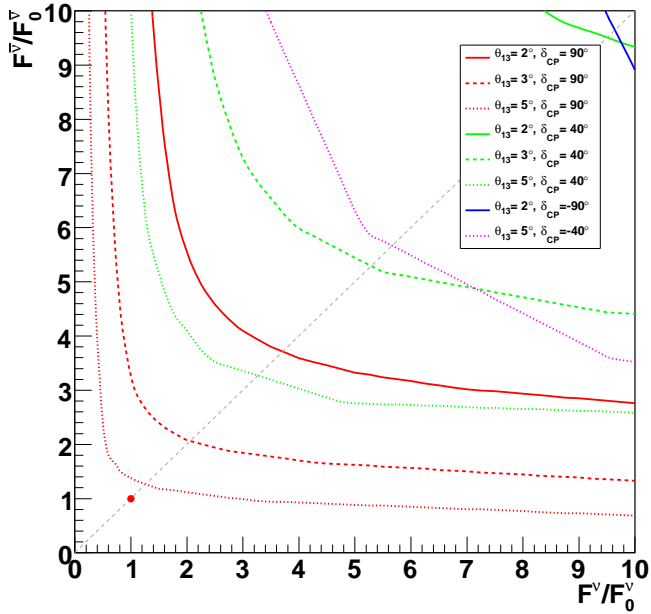} }
\caption{CP discovery potential as a function of both neutrino and antineutrino fluxes ($F$ and $\bar F$, respectively) with respect to the nominal flux $F_0$, for several input values of $\delta$ and $\theta_{13}$, shown in the legend. The dashed black line determines the points were the same neutrino and antineutrino fluxes are considered, $F = \bar F$. The red circle shows the point where both fluxes correspond to the reference value, $F_0 = 3\times 10^{18}$ useful ion decays per year. Lines of the same color correspond to the same $\delta$ values, while lines of the same type (continuous, dotted, dashed) correspond to the same values of $\theta_{13}$. For the points in the region above and to the right of each line, CP violation can be established at $3\sigma$ 1 d.o.f. after marginalizing over the rest of parameters, at least for one point in the parameter space.} 
 \label{fig:cp_nuflux_nubarflux}
 \end{center}
\end{figure}

In Fig.~\ref{fig:cp_nuflux_nubarflux}, the discovery potential is
presented as a function of the neutrino and antineutrino fluxes $F$
and $\bar F$ with respect to $F_0 $, for several representative values
of $\delta$ and $\theta_{13}$. For the points in the region below and
to the left of each line, CP violation cannot be established at
$3\sigma$ 1 d.o.f. after marginalizing over the rest of
parameters. The diagonal dashed black line represents the points where
the same neutrino and antineutrino flux ratios are considered, $F =
\bar F$. Notice that, if we restrict both fluxes to the reference
value $F_0$ (marked as a red circle in the plot), then CP violation
cannot be determined for any pair of the ($\delta$,$\theta_{13}$)
values we have considered. Even for maximal CP violation ($\delta=\pm
90^\circ$) and increasing $\theta_{13}$ up to $5^\circ$, a 40\%
increase of the flux for both polarities is needed in order to
establish CP violation at least for one point in the parameter space.
On the other hand, if we fix the neutrino flux ratio at $F/F_0=1$, 
maximal CP violation can be established at 3$\sigma$ for $\theta_{13} = 3^\circ (5^\circ)$
if we manage to achieve an antineutrino flux ratio $\bar F/F_0=3 (1.4)$. 
It is also important to notice that,
for negative values of $\delta$ only two of the lines are visible in
the plot, corresponding to the input values ($\delta=-90^\circ$,
$\theta_{13}=5^\circ$) and ($\delta=-40^\circ$,
$\theta_{13}=2^\circ$), but no line is present in the plot for
$\theta_{13}=3^\circ$ and $\delta<0$. This is due to the so-called
``$\pi$-transit'' effect \cite{Huber:2002mx}: matter effects mimic
true CP violation and, for this particular value of $\theta_{13}$,
when $\delta<0$ the so-called ``sign clones''\footnote{Degenerate minima of the $\chi^2$ corresponding to 
a wrong assignment of the neutrino mass hierarchy and to a different pair ($\theta_{13},\delta$), 
see Refs.~\cite{Minakata:2001qm,Donini:2003vz,Minakata:2010zn}.}
move from the true CP-violating values to CP-conserving ones. 
As a consequence, in this particular region of
the parameter space CP violation cannot be established even if it is
maximal\footnote{This effect depends strongly on the amount of matter
effect observed at the considered setup. For the Neutrino Factory,
where it was discussed first, the $\pi$-transit occurs for $\sin^2 2
\theta_{13} \sim 10^{-3} (\theta_{13} \sim 1^\circ)$.}.

\begin{figure}
\begin{center}
\includegraphics[width=0.6\textwidth]{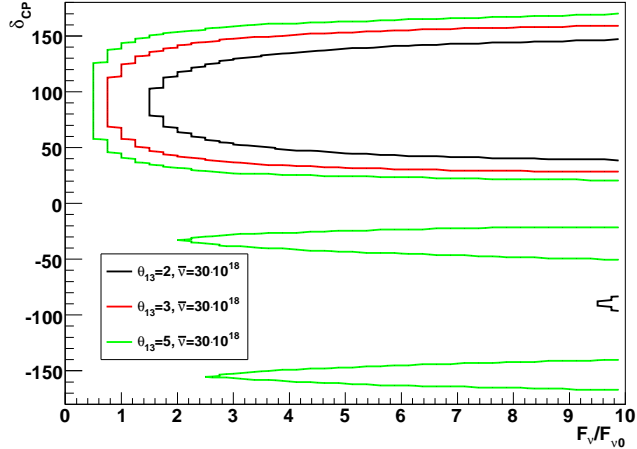} 
\caption{CP discovery potential as a function of the neutrino flux $F$ and $\delta$, for several values of $\theta_{13}$, and an antineutrino flux ratio $\bar F/F_0=10$. For the points in the region to the left of each line, CP violation cannot be established at $3\sigma$ 1 d.o.f. after marginalizing over the rest of parameters.} 
\label{fig:cp_nuflux}
\end{center}
\end{figure}

This effect can also be appreciated in Fig.~\ref{fig:cp_nuflux}, where the CP discovery potential is plotted as a function of the neutrino flux ratio, $F/F_0$, for $\theta_{13}=2^\circ,\,3^\circ,$ and $5^\circ$, keeping the antineutrino flux ratio fixed at $\bar F/F_0=10$. For the points located to the left of each line in the plot, CP violation cannot be established at $3\sigma$ CL after marginalization over the rest of parameters. It can also be seen here that the CP discovery potential is quite poor in the $\delta<0$ region: for $\theta_{13}=2^\circ$, we are only sensitive to CP violation if it is maximal for extremely high values of the neutrino flux ratio, while for the $\theta_{13}=5^\circ$ case two narrow bands appear around $\delta=-40^\circ$ and $-150^\circ$. Again, when $\theta_{13}=3^\circ$, CP violation cannot be established for any negative value of $\delta$. 

\begin{figure}
\begin{center}
{\includegraphics[width=0.6\textwidth]{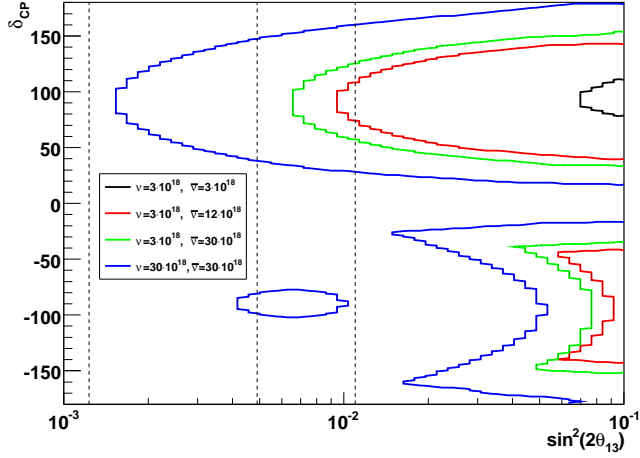} }
\caption{CP discovery potential as a function of $\sin^2(2 \theta_{13})$ and $\delta$, for several values of both neutrino and antineutrino flux ratios, as indicated in the legend. The dashed vertical lines show, from left to right, the values of $\sin^2(2\theta_{13})$ corresponding to $\theta_{13}=1^\circ,\,2^\circ$ and $3^\circ$, respectively. For the points in the region to the left of each curve, CP violation cannot be established at $3\sigma$ 1 d.o.f. after marginalizing over the rest of parameters.} 
 \label{fig:cp_theta13_delta}
 \end{center}
\end{figure}

Finally, we show in Fig.~\ref{fig:cp_theta13_delta} the CP discovery
potential as a function of $\theta_{13}$ and $\delta$, for several
values of the neutrino and antineutrino flux ratios. Notice the
vertical dotted lines, which indicate, from left to right in the plot,
the values of $\sin^2(2\theta_{13})$ corresponding to
$\theta_{13}=1^\circ,\,2^\circ$ and $3^\circ$, respectively. We see
again a strong lack of sensitivity around $\sin^2(2 \theta_{13})\sim
10^{-2}$, which corresponds to $\theta_{13}\sim 3^\circ$, due to the
$\pi$-transit phenomenon.  It can also be seen how, for smaller values
of $\theta_{13}$, we recover some sensitivity to CP violation.

Regarding the statistical dependence of the setup, a strong improvement takes place when the antineutrino flux is increased from $ F_0 \rightarrow 4 F_0$, even though we keep the neutrino flux fixed at $F_0$. However, once we have reached this point, we get practically no improvement at all if we keep increasing the antineutrino flux unless the neutrino flux is also enhanced. This can be seen from the comparison of the red and green lines: we have increased the antineutrino flux another factor 2.5 (up to $10 F_0$), but the CP discovery potential improvement is quite mild. This is due to the fact that, in order to achieve sensitivity to the CP-violating phase, a comparison between the neutrino and antineutrino oscillated events is mandatory: even if we continue increasing the antineutrino flux the CP discovery potential will not increase unless we have enough neutrino events to compare with. This is precisely what happens when we compare the green and blue lines in the plot: the improvement is remarkable in this case, though only the neutrino flux has been enhanced, because now all the antineutrino events are useful.

\subsection{Sensitivity to the neutrino mass hierarchy}

\begin{figure}[t!]
\vspace{-0.5cm}
\begin{center}
\begin{tabular}{cc}
\hspace{-0.3cm} \epsfxsize8cm\epsffile{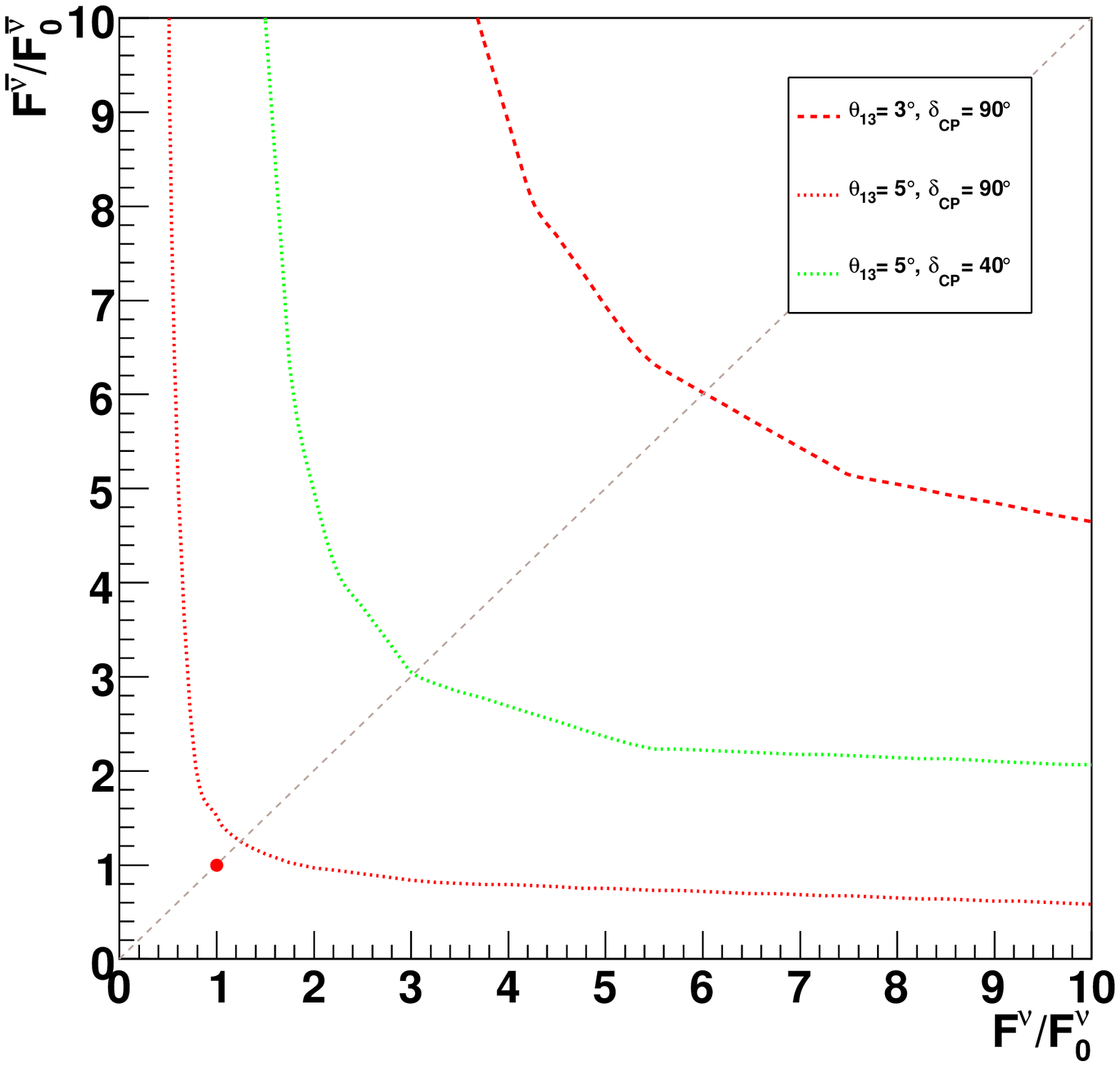} &
\hspace{-0.3cm} \epsfxsize8cm\epsffile{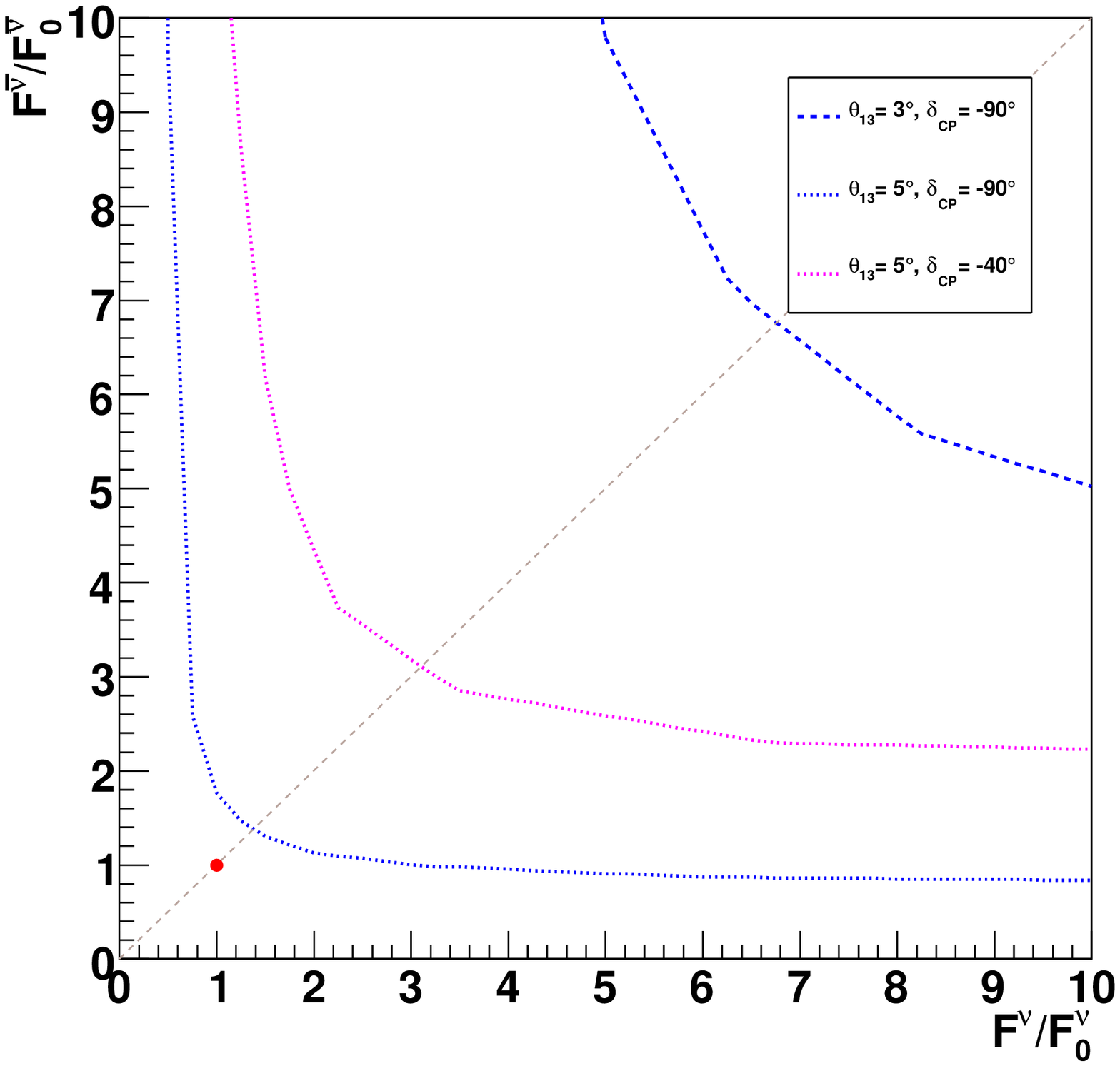} 
\end{tabular}
\caption{Sensitivity to the hierarchy as a function of both neutrino and antineutrino fluxes ($F$ and $\bar F$, respectively) with respect to the nominal flux $F_0$, for several input values of $\delta$ and $\theta_{13}$, shown in the legend. Left panel: normal hierarchy; right panel: inverted hierarchy.
The dashed black line determines the points were the same neutrino and antineutrino fluxes are considered, $F = \bar F$. The red circle shows the point where both fluxes correspond to the reference value, $F_0 = 3\times 10^{18}$ useful ion decays per year. Lines of the same color correspond to the same $\delta$ values. For the points in the region above and to the right of each line, a given hierarchy can be established at $3\sigma$ 1 d.o.f. after marginalizing over the rest of parameters, for the considered point in the parameter space.}
\label{fig:hierarchy_nuflux_nubarflux}
\end{center}
\end{figure}

In Fig.~\ref{fig:hierarchy_nuflux_nubarflux}, the sensitivity to the neutrino mass hierarchy is presented as a function of the neutrino and antineutrino fluxes $F$ and $\bar F$ with respect to $F_0 $, for several representative values of $\delta$ and $\theta_{13}$. The left panel refers to the normal hierarchy, the right panel to the inverted one. For the points in the region above and to the right of each line, a given hierarchy can be established at $3\sigma$ 1 d.o.f. after marginalizing over the rest of parameters, for the particular choice of input parameters. 
The diagonal dashed black line represents the points where the same neutrino and antineutrino flux ratios are considered, $F = \bar F$. Notice that, if we restrict both fluxes to the reference value $F_0$ (marked as a red circle in the plot), then the hierarchy cannot be determined for any pair of the ($\delta$,$\theta_{13}$) values we have considered. Also in this case, 
as it was for the CP violation discovery potential, we need  a 40\% increase of the flux for both polarities in order to establish a given hierarchy at least for the largest considered $\theta_{13}$ value, $\theta_{13} = 5^\circ$. As we have said, neutrino fluxes are unlikely to be much higher than the reference value. However, if we fix the neutrino flux ratio at 
$F/F_0=1$, with a 50\% increase of the antineutrino flux we become sensitive to the hierarchy for this particular point in the parameter space, $\theta_{13} = 5^\circ$, $\delta = + 90^\circ$ 
or $-90^\circ$ (for normal or inverted hierarchy, respectively).
For smaller $\theta_{13}$, we need a neutrino flux $F = 2 F_0$, at least (if, at the same time, we manage that the antineutrino flux is increased to $\bar F \sim 4 F_0$). 

\begin{figure}
\begin{center}
{\includegraphics[width=0.7\textwidth]{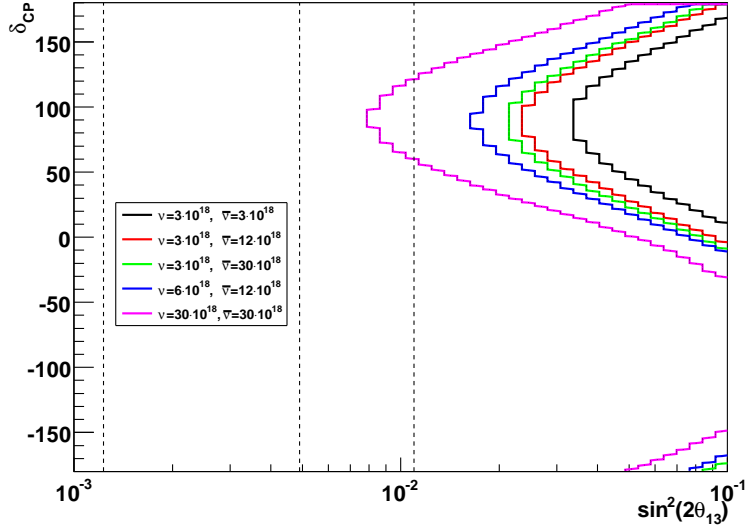} }
\caption{ The $\textrm{sgn}(\Delta m_{31}^2) $ sensitivity as a function of  $\sin^2(2\theta_{13})$ and $\delta$ for several values of both neutrino and antineutrino flux ratios, assuming normal hierarchy. The dashed vertical lines indicate, from left to right, the values of $\sin^2(2\theta_{13})$ corresponding to $\theta_{13}=1^\circ,\,2^\circ$ and $3^\circ$, respectively. For the points in the region to the left of each curve, the correct hierarchy cannot be established at $3\sigma$ 1 d.o.f. after marginalizing over the rest of parameters. } 
 \label{fig:NH}
 \end{center}
\end{figure}

\begin{figure}
\begin{center}
{\includegraphics[width=0.7\textwidth]{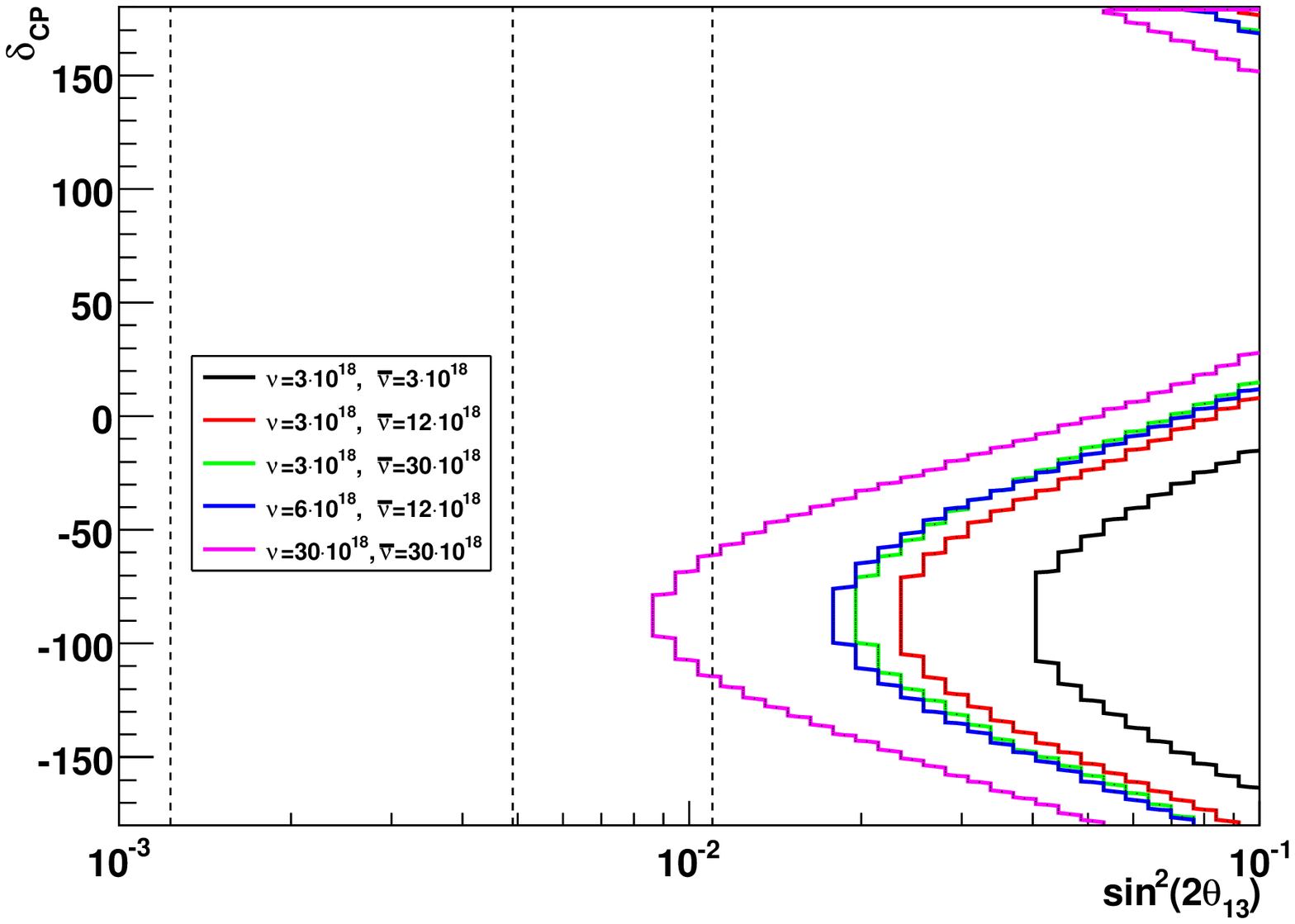} }
\caption{The $\textrm{sgn}(\Delta m_{31}^2) $ sensitivity as a
function of $\sin^2(2 \theta_{13})$ and $\delta$ for several values of
both neutrino and antineutrino flux ratios, assuming inverted
hierarchy. The dashed vertical lines indicate, from left to right, the
values of $\sin^2(2\theta_{13})$ corresponding to
$\theta_{13}=1^\circ,\,2^\circ$ and $3^\circ$, respectively. For the
points in the region to the left of each curve, the correct hierarchy
cannot be established at $3\sigma$ 1 d.o.f. after marginalizing over
the rest of parameters.}
 \label{fig:IH}
 \end{center}
\end{figure}

As we have already mentioned in Sec.~\ref{subsec:CP}, as the baseline of the setup is relatively ``short'', matter effects turn out to be quite mild and therefore we are sensitive to the mass hierarchy only in a small region of the parameter space. The sensitivity to $\textrm{sgn}(\Delta m_{23}^2)$ is depicted in Fig.~\ref{fig:NH} (Fig.~\ref{fig:IH}) as a function of $\sin^2(2\theta_{13})$ and $\delta$, assuming normal (inverted) hierarchy, for several values of the neutrino and antineutrino fluxes. The vertical dashed lines indicate, from left to right, the values of $\sin^2(2\theta_{13})$ corresponding to $\theta_{13}=1^\circ,\, 2^\circ $ and $3^\circ $, respectively. For the points located to the left of each curve, the correct hierarchy cannot be determined at a statistical significance of $3\sigma$, after marginalization over the rest of parameters.  

It can be seen that changing from normal to inverted hierarchy is practically equivalent to replacing $\delta\rightarrow -\delta$. In the limit of null matter effect, the sensitivity to $\textrm{sgn}(\Delta m_{31}^2) $ comes from the CP-violating term in the probability, which for normal hierarchy is maximal for neutrinos for positive values of $\delta$, while for the inverted hierarchy it is maximal for antineutrinos and $\delta<0$. The small  asymmetry observed when we change from normal to inverted hierarchy and \emph{viceversa} is due to a combination of two factors: on one side, matter effects enhance neutrino with respect to antineutrino events when the hierarchy is normal, while the opposite effect takes place if the hierarchy is inverted. On the other hand, as the antineutrino cross section is smaller,  the lines with the same fluxes for both polarities are, in general, worse if we assume inverted hierarchy than if we assume normal hierarchy.

\begin{figure}
\begin{center}
{\includegraphics[width=0.8\textwidth]{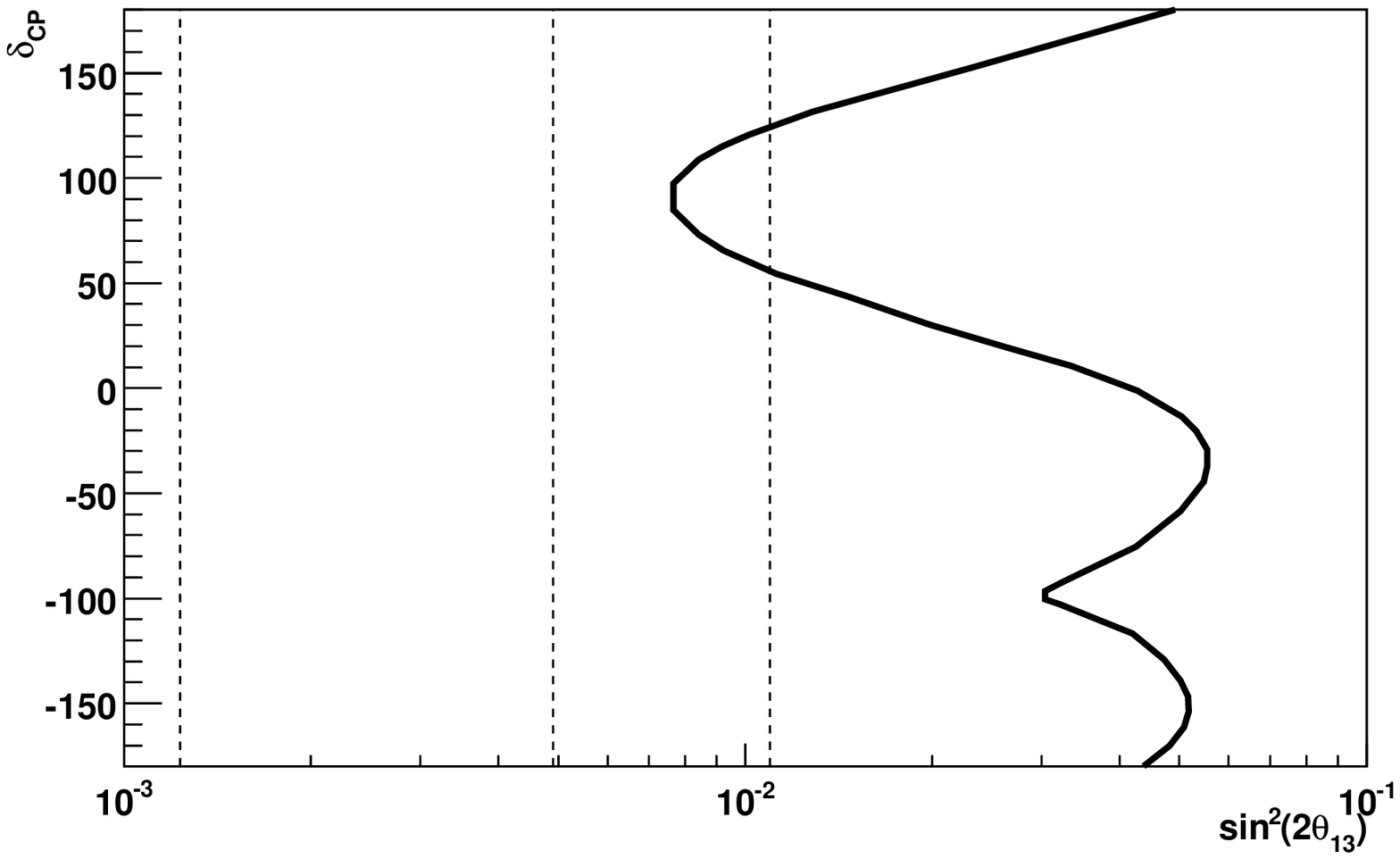} }
\caption{ The $\textrm{sgn}(\Delta m_{31}^2) $ sensitivity as a
function of $\sin^2(2\theta_{13})$ and $\delta$ for $3 \times 10^{19}$
decays per year both for neutrinos and antineutrinos, assuming normal
hierarchy. The dashed vertical lines indicate, from left to right, the
values of $\sin^2(2\theta_{13})$ corresponding to
$\theta_{13}=1^\circ,\,2^\circ$ and $3^\circ$, respectively. For the
points in the region to the left of each curve, the correct hierarchy
cannot be established at $3\sigma$ 1 d.o.f. after marginalizing over
the rest of parameters. }
 \label{fig:atm}
 \end{center}
\end{figure}

As noted in Ref.~\cite{Donini:2007qt}, a magnetized iron detector can
fruitfully combine data from the Beta Beam and from atmospheric
neutrinos to improve the sensitivity on the mass
hierarchy~\cite{TabarellideFatis:2002ni}. For large values of
\thetaot, such combination can be of value for the present setup in the
occurrence of normal (inverted) hierarchy and negative (positive)
values of $\delta$, i.e. in the region of null sensitivity of
Fig.~\ref{fig:NH} and \ref{fig:IH}. The combination is depicted in
Fig.~\ref{fig:atm} assuming $F = \bar F = 10 F_0$.  In
particular, for $\delta=-90^\circ$ atmospheric neutrinos bring the
3$\sigma$ sign sensitivity of the setup down to $\sin^2 2\theta_{13}
\simeq 3 \times 10^{-2}$.

\section{Conclusions}
\label{sec:conclusions}

Since its inceptions, Beta Beams have been conceived to exploit in an
optimal manner existing facilities in order to establish CP violation
in the leptonic sector.  In this paper, we considered a setup that
leverages at most present European infrastructures. It is based on the
CERN-SPS accelerator, which is employed to boost high-Q ions toward
the Hall C of the Gran Sasso Laboratories. 

For a far detector of
100~kton mass, a $\beta^+$-emitters ($^8$B) flux of approximately $6 \times 10^{18}$ useful
decays per year\footnote{This is about three times the flux proposed for $^{18}$Ne, where $F_0 \sim 2 \times 10^{18}$. } is needed to observe CP violation in a large fraction
of the parameter space (60\%) for any value of $\theta_{13}$ that gives a positive signal at T2K ($\theta_{13} \simge 3^\circ$).
This sensitivity to $\delta$ is deteriorated for $\delta<0$ due to the
occurrence of the $\pi$-transit, as observed in other facilities.
The $^8$B flux  must be accompanied by a $^8 Li$ flux of $\sim 3\times 10^{19}$ decays per year. 
Present studies on the ionization cooling technique or on ISOL-type targets indicate that such a large $^8$Li flux could be feasible.
Moreover, the former technique should produce $\beta^+$ and $\beta^-$ emitters at a similar rate  although $^8$B ions interact stronger than $^8$Li ions with 
materials in the target and in the recollection region.  To achieve the fluxes above clearly represents the most challenging task for accelerator R\&D
but it is a viable option with respect to  $^{18}$Ne, where  ISOL-type targets fall almost two orders of magnitude short
of the goal. 

In the same configuration, we find a non-negligible sensitivity to the neutrino mass hierarchy that extends
up to $\theta_{13} \simeq 4^\circ$ for positive (negative) values of $\delta$ for normal (inverted) hierarchy. 
In the opposite parameter area, i.e. for negative (positive) values of $\delta$ and inverted
(normal) hierarchy, the combination with atmospheric data collected
during the Beta Beam run by the same magnetized detector further
improves such sensitivity at large \thetaot ($\simeq 6^\circ$).
Combination of atmospheric data with Beta Beam--driven ones should also be able to solve part of the $\pi$-transit deterioration 
discussed above.

\section*{Acknowledgments}

We wish to express our gratitude to T. Tabarelli de Fatis for many
useful information on the simulation of magnetized iron detector. We
gratefully acknowledge M. Mezzetto and F. Ronga for several interesting
discussions.
This work was supported by the European Union under the European Commission
Framework Programme 07 Design Study "EURO$\nu$", project 212372. P.C. and A.D. acknowledge funding by the spanish 
ministry for Science and Innovation under the project FPA2009-09017 and the program "CUP" Consolider-Ingenio 2010, project CSD2008-0037; by the Comunidad Aut\'onoma 
de Madrid through project HEPHACOS-CM (S2009ESP-1473). Eventually, P.C. acknowledges financial support from the Comunidad Aut\'onoma de Madrid.


\end{document}